\documentclass{emulateapj}
\slugcomment{Draft Version, Accepted for publication in ApJL}
\usepackage{times}
\usepackage{bm}

\newcommand{\bolB}{{\bm  B}}

\newcommand{\beq}{\begin{equation}}
\newcommand{\eeq}{\end{equation}}
\newcommand{\beqn}{\begin{eqnarray}}
\newcommand{\eeqn}{\end{eqnarray}}
\newcommand{\beqno}{\begin{equation*}}
\newcommand{\eeqno}{\end{equation*}}
\newcommand{\beqnno}{\begin{eqnarray*}}
\newcommand{\eeqnno}{\end{eqnarray*}}

\shorttitle{Structure of the M87 Jet}
\shortauthors{Asada \& Nakamura}

\begin{document}

\title{The Structure of the M87 Jet: A Transition from Parabolic to Conical Streamlines}

\author{Keiichi Asada\altaffilmark{1} and Masanori Nakamura\altaffilmark{1,2,3}}
\affil{
$^1$Institute of Astronomy \& Astrophysics, Academia Sinica, P.O. Box
23-141, Taipei 10617, Taiwan \\
{\tt asada@asiaa.sinica.edu.tw, nakamura@asiaa.sinica.edu.tw}}
\affil{
$^2$Department of Physics and Astronomy, The Johns Hopkins University, 3400 N. Charles Street,
Baltimore, MD 21218, USA}
\affil{
$^3$Space Telescope Science Institute, 3700 San Martin Drive, Baltimore,
MD 21218, USA}

\begin{abstract}
The structure of the M87  jet, from milli-arcsec to arcsecond scales, is
extensively  investigated,  utilizing the  images  taken  with the  EVN,
MERLIN  and  VLBA.  We  discover  that  the  jet maintains  a  parabolic
streamline  over  a range  in  sizescale  equal  to $10^{5}$  times  the
Schwarzschild  radius.  The jet  then transitions  into a  conical shape
further downstream.  This suggests  that the magnetohydrodynamic  jet is
initially  subjected to  the confinement  by the  external gas  which is
dominated by the gravitational influence of the supermassive black hole.
Afterwards the jet is then  freely expanding with a conical shape.  This
geometrical transition  indicates that the  origin of the  HST-1 complex
may  be a consequence  of the  over-collimation of  the jet.  Our result
suggests  that when  even higher  angular  resolution is  provided by  a
future submm VLBI  experiment, we will be able to  explore the origin of
active galactic nuclei jets.
\end{abstract}


\section{Introduction}

M 87  is one  of the nearest  active galaxies \citep[16.7  Mpc; ][]{J05}
which  exhibits   relativistic  outflows.   The  mass   of  the  central
supermassive black  hole (SMBH) is  measured to be  3.2 $\times$10$^{9}$
$M_{\sun}$  from HST observations  of the  ionized gas  disk \citep[{\em
e.g.},][]{M97}.   Recent analyses  of the  stellar kinematics  suggest a
larger  mass of  6.6 $\times$  10$^{9}$ $M_{\sun}$  \citep[]{G11}.  This
larger  mass   gives  an  apparent   size  $\sim$  8  $\mu$as   for  the
Schwarzschild radius  $r_{s}$.  This galaxy therefore  provides a unique
opportunity to study the  relativistic outflow with the highest physical
resolution in units of r$_{s}$.

Based on  the VLA/VLBI observations  during the past three  decades, the
structure of  the M87 jet  has been extensively  examined. \citet[]{J99}
found a  smooth variation  of the jet  opening angle from  60$^\circ$ at
$\sim$ 0.03 pc to smaller  than 10$^\circ$ over $\sim$ 10 pc, indicating
that  the  jet is  being  collimated  by  the magnetohydrodynamic  (MHD)
process \citep[]{MKU01}.   One of the  most interesting features  in the
M87 jet is the bright knot G, lying about 1$\arcsec$ ($\sim$ 78 pc) from
the core in the VLA maps \citep[]{O89}. That region has been resolved by
the HST into  a structured complex known as  ``HST-1'' \citep[]{B99}. It
is located around 0.8 - 1.0$\arcsec$ (or $62 - 78$ pc) from the core. It
consists  of   bright  knots,   whose  (apparent)  proper   motions  are
superluminal  with  a range  of  $4c  -  6c$ \citep[]{B99}.   A  similar
velocity pattern has been observed  at radio frequencies using the VLBA.
The component  of HST-1  which is furthest  upstream (i.e.,  HST-1d), is
stationary to within  the errors ($<0.25c$), and has  been identified as
the origin  of the superluminal ejections  \citep[]{C07}.  The structure
of the jet downstream of HST-1 ($1 - 18 \arcsec$ or $0.1 - 1.5$ kpc) can
be  characterized by  a conical  shape with  an opening  angle  of $\sim
6^{\circ}$  \citep[]{O89}.  HST-1 is  also a  remarkable site  for large
flaring activity across the  electromagnetic spectrum from radio through
optical to  X-ray bands  \citep[]{M09, H09}.  Chandra  observations show
that the  flaring event started  in 2000, and its  brightness eventually
increased more than  50 times.  Peak brightness occurred  in 2005 and an
adiabatic  compression  has been  suggested  as  a  cause of  the  flare
\citep[]{H06}.   It  is  thought  to  be  one  possible  sites  for  TeV
$\gamma$-ray  emission \citep[]{H09}.   Note that  all distances  are in
projection throughout this section.

\section{Observations and Data Reductions}

\subsection{EVN data}
We conducted  the EVN  observations towards M  87 on  7 March 2009  at a
wavelength of  18 cm with Cambridge (UK),  Effelsberg (Germany), Jodrell
Bank  (UK),  Knockin  (UK),   Medicina  (Italy),  Noto  (Italy),  Onsala
(Sweden), Torun  (Poland), and Westerbork  (Netherlands) stations.  Both
left  and  right  circular  polarization  data  were  recorded  at  each
telescope using  8 channels of 8  MHz bandwidth and 2  bit sampling. The
data were correlated at the JIVE correlator.

{\it A priori} amplitude calibration for each station was derived from a
measurement of the  system temperatures during each run  and the antenna
gain.  Fringe  fitting was performed  using AIPS.  After delay  and rate
solutions were  determined, the  data were averaged  over 12  seconds in
each IF and self-calibrated using Difmap.

\subsection{MERLIN data}
We conducted the MERLIN observations on  9 March 2007 at a wavelength of
18  cm with  Defford,  Cambridge, Knockin,  Darnhall,  Jodrell Bank  and
Tabley stations.   Both left and  right circular polarization  data were
transferred from each telescope with 15 MHz bandwidth.

Initially, the data were  pipelined, and a channel-by-channel flux scale
derived from 3C 286 observation  was applied.  After this {\it a priori}
amplitude calibration, we self-calibrated the data using Difmap.

\subsection{VLBA data}
We also  analyzed archival  VLBA data  of the M  87 jet  (BK073).  Those
observations were carried out on 22  January 2000 at 15 GHz with all ten
stations of the  VLBA and one station  of the VLA.  4 IFs,  each with an
8-MHz bandwidth,  were recorded at  each telescope.  Data  reduction was
conducted in the same manner as for the EVN observations.  We reproduced
an M 87 image, which is consistent with that published in \citet[]{K07}.

\begin{figure}
\begin{center}
\includegraphics[scale=1.0,  angle=0]{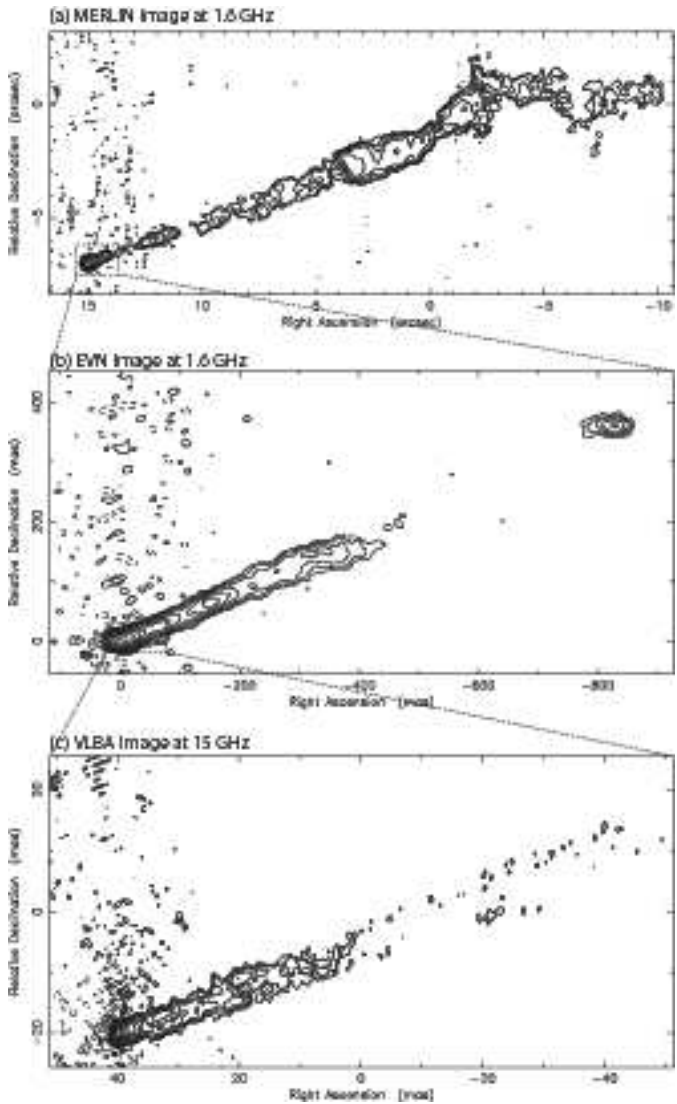} 
\caption{\label{IPOL-maps}
(a) MERLIN image of  M 87.  Contours are plotted at -1,  1, 2, ..., 1024
$\times$  3.64  mJy  beam$^{-1}$,  which  is  three-times  the  residual
r.m.s. noise.  The  image is restored with a circular  beam of 25 arcsec
to emphasize the  extended structure.  (b) EVN image  of M 87.  Contours
are plotted at -1, 1, 2,  ..., 1024 $\times$ 1.37 mJy beam$^{-1}$, which
is three-times the residual r.m.s.  noise.  The synthesized beam is 17.8
mas  $\times$ 13.1  mas  with the  major  axis at  a  position angle  of
76.1$^{\circ}$.  (c) VLBA image of M 87.  Contours are plotted at -1, 1,
2, ...,  1024 $\times$  0.45 mJy beam$^{-1}$,  which is  three-times the
residual r.m.s. noise.   The synthesized beam is 1.22  mas $\times$ 0.59
mas with the major axis at a position angle of -3.99$^{\circ}$.}
\end{center}
\end{figure}

\section{Results}
We  show  the   images  from  EVN,  MERLIN  and   VLBA  observations  in
Fig. \ref{IPOL-maps}.   Image qualities are summarized in  table 1.  The
bright core at  the eastern edge of the jet  and continuous jet emission
are detected  in all images.  We  define the core with  a Gaussian model
which was fitted to the innermost  bright region in all the images.  The
isolated component at  the distance of 900 mas from the  core in the EVN
image is the  HST-1 knot.  It corresponds to  the first bright component
to  the west of  the core  in the  MERLIN image.   This feature  was not
detected in the VLBA image at  15 GHz, since that data were taken before
the  HST-1 component  underwent  its  flaring event.   One  of the  most
important  achievements in  our  observations is  a  clear detection  of
continuous jet  emission up to 500 mas  from the core in  the EVN image.
The continuous jet structure between  160 and 500 mas has been suggested
in previous  measurements \citep[]{R89,C07}, but here  they are robustly
detected and resolved with high significance in our EVN observations.

In order to illustrate the structure of the M 87 jet, we analyze the jet
emission  along a  position angle  of  338$^{\circ}$ and  fit the  cross
sections of the jet with one  Gaussian or multiple Gaussians for all the
images.  We define the width of  the jet as the deconvolved FWHM for the
one-Gaussian fitting  case.  For the multiple-Gaussian  fitting case, we
define the distance between the outer edges of the FWHMs as the width of
the jet.  We note that the  jet cross sections can be represented mainly
by  two  Gaussians  when   multiple  Gaussians  were  fitted.   This  is
consistent  with  the  edge   brightened  model  suggested  by  previous
observations \citep[]{O89,R89,J99}.

\begin{figure*}
\centerline{{\includegraphics[scale=1.35, angle=0]{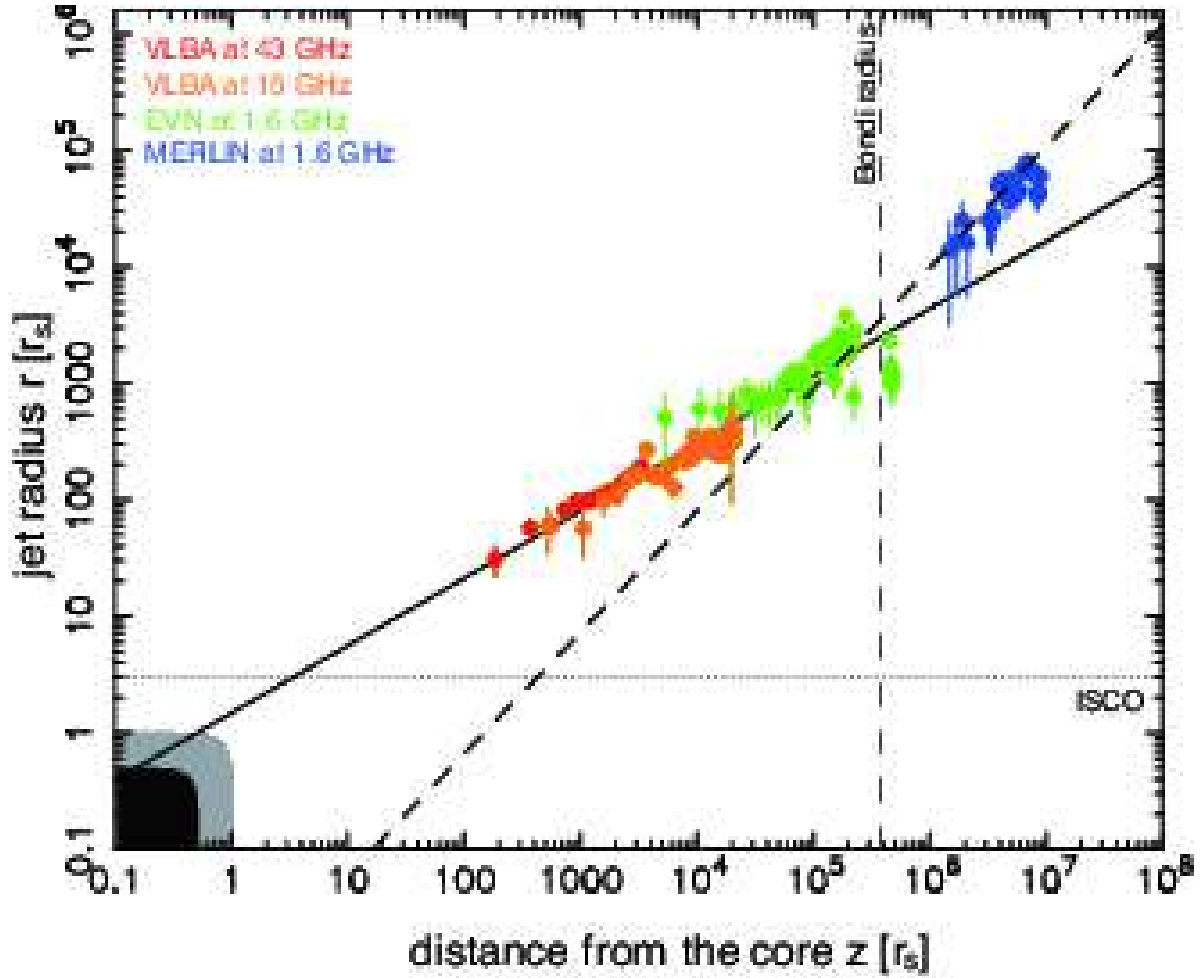}}}
\caption{\label{r-z}
Distribution of the  radius of the jet as a  function of the deprojected
distance from the core in units  of $r_{s}$.  We used images obtained by
previous VLBA measurements at 43 GHz (red circles) and at 15 GHz (orange
circles),  EVN  measurements  at  1.6  GHz (green  circles)  and  MERLIN
measurements at  1.6 GHz  (blue circles).  The  jet is described  by two
different shapes.  The solid line indicates a parabolic structure with a
power law  index $a$ of 1.7,  while the dashed line  indicates a conical
structure with $a$ of 1.0.   HST-1 is located around 5 $\times$ 10$^{5}$
$r_{s}$.   The black  area shows  the size  of minor  axis of  the event
horizon of  the spinning  black hole with  maximum spin.  The  grey area
indicates  the size  of  the major  axis  of the  event  horizon of  the
spinning black  hole with maximum spin,  and corresponds to  the size of
the  event horizon  of the  Schwarzschild black  hole.  The  dotted line
indicates  the size of  the inner  stable circular  orbit (ISCO)  of the
accretion disk for the Schwarzschild black hole.}
\end{figure*}

We show the radius of the  jet as a function of the deprojected distance
from the  core in  units of  $r_{s}$ in Fig.  \ref{r-z}.  Note  that the
radius  as  defined  perpendicular  to  the jet  does  not  suffer  from
orientation effects.   We assume  that the viewing  angle of the  jet is
14$^{\circ}$ based on a  beaming analysis \citep[]{WZ09}.  Hereafter, we
use the  deprojected distance  along the jet  throughout the  paper.  We
note that 1 mas corresponds  to 530 $r_{s}$ in deprojected distance.  We
measure  the radius  of the  jet at  the innermost  region by  using the
previous VLBA image at 43 GHz  by \citet[]{W08} as well.  The jet radius
increases downstream, as clearly shown in Fig. \ref{r-z}.  We found that
the  streamlines  of the  M87  jet can  be  divided  into two  different
regimes: i)  a parabolic shape  with $a$ =  1.73 $\pm$ 0.05  ($z \propto
r^{a}$, where $r$ is the radius of the jet emission and $z$ is the axial
distance from  the core), on scales  up to $\sim$  2.5 $\times$ 10$^{5}$
$r_{s}$, and ii)  a conical shape with $a$ = 0.96  $\pm$ 0.1 starting at
around 2.5  $\times$ 10$^{5}$  $r_{s}$.  It is  remarkable that  the jet
propagates  with a  single  power-law streamline  over  more than  three
orders of  magnitude in distance.   Furthermore, it is notable  that the
two   observed   power-law   streamlines   intersect   at   2   $\times$
10$^{5}\,r_{s}$,  which is  close to  the Bondi  radius of  3.8 $\times$
10$^{5}$ $r_{s}$ ($\sim$ 250  pc\footnote{The original value is replaced
by taking into account of  the mass difference of SMBH.}) \citep[]{A06}.
Indeed, this is the  first observational evidence detecting a transition
from parabolic to conical streamlines in extragalactic jet systems.

\section{Discussion and Summary}
\label{sec:Discussion-Conclusions}

\subsection{Unconfined Structure: Downstream of HST-1}

We  consider  a  conical streamline  ($z  \propto  r^{a},  a =  1$)  for
supersonic  jets. In  an adiabatic  jet, the  internal  pressure $p_{\rm
jet}$ decreases with axial distance $z$ as $z^{-2\Gamma}$ ($\Gamma$: the
ratio of specific heats).  So,  we speculate that the constant expansion
of the jet radius and a  conical structure downstream of HST-1 to knot A
in the  M87 jet, requires the  same axial gradient for  the external ISM
pressure,  $p_{\rm ism}$,  as $p_{\rm  jet}$ \citep[]{O89}.   If $p_{\rm
ism}$ decreases  slower than $p_{\rm  jet}$, then $p_{\rm ism}  > p_{\rm
jet}$ at some  distance so that a recollimation  shock will be triggered
\citep[]{S83}.  The  self-similar solution  of a conical  streamline for
the magnetized  case (with a  purely toroidal field  component) requires
$p_{\rm  ism}  \propto  z^{-b},   b=4$  \citep[]{Z08}.   For  a  general
(non-self-similar) case, $b > 2$  is allowed in analytical and numerical
models  \citep[]{T08,  L09,  K09}.  X-ray observations  reveal  the  ISM
properties such as the Bondi radius $r_{\rm B} \sim$ 250 pc and the King
core radius $r_{\rm c} \simeq 1.4$ kpc \citep[see, {\em e.g.},][]{YWM02,
D03, A06}.  Thus, the region of  conical streamlines in the M87 jet lies
between  $r_{\rm B}$  and the  marginal radius  for the  power-law decay
beyond  $r_{\rm  c}$  in  the  King profile,  indicating  that  the  ISM
distribution  is  essentially  uniform.   We  thus  rule  out  that  the
structure  downstream of  HST-1 is  hydrostatically confined  by $p_{\rm
ism}$ in order to conform to a conical streamline.

In order  to possess a  conical streamline without  any over-collimation
between knots  HST-1 and A, the  condition $ p_{\rm  jet} \gtrsim p_{\rm
ism}$  should  be  maintained.   Knots  HST-1  to  A  do  appear  to  be
over-pressured    with     respect    to    the     external    pressure
\citep[]{O89}.  However, $p_{\rm  jet}$  of the  inter-knot regions,  as
estimated by the minimum energy argument for the VLA data \citep[]{S96},
appears under-pressured  with respect to $p_{\rm ism}$,  as estimated by
the recent X-ray observations  \citep[]{YWM02, R06}.  One possibility is
an  underestimation of  the magnetic  field strength  when  the toroidal
(azimuthal) components  are not  considered, cf.  \citet[]{O89}.  It has
been further  suggested that  the magnetic field  energy is at  least in
equipartition (or  even larger  with a  factor of 1  $\sim$ 2)  with the
energy of the radiating ultrarelativistic electrons \citep[]{S05}.

We note that over-pressured knots do appear to have trails of stationary
recollimation  shocks  in   purely  hydrodynamic  jets.   \citet[]{FW85}
performed  hydrodynamic simulations  to  apply stationary  recollimation
shocks to the  observed knots at VLA scales under  the assumption of the
shallow   ISM  gradient   ($p_{\rm  ism}\propto   z^{-1}$).   Stationary
features,  however,  are in  conflict  with  the  observed large  proper
motions \citep[]{B95, B99},  while the ISM also does  not appear to have
such  a gradient.   Therefore,  we suggest  that  the highly  magnetized
nature of  the jet may  be responsible for  the conical part of  the M87
jet.

\subsection{Confined Structure: Upstream of HST-1}

We next consider  a parabolic streamline ($1 < a  \le 2$) for supersonic
jets.   It  is  shown  that  the  magnetized jet  can  be  parabolic  in
analytical and numerical models where  the ISM pressure is decreasing as
$p_{\rm ism}  \propto z^{-b},  b=2$ \citep[]{T08, K09}.   A self-similar
solution,  also exists for  non-magnetized cases  with a  pure parabolic
streamline ($a=2$) under the same  $p_{\rm ism}$ dependence with $b = 2$
\citep[]{Z08}.   Their solution indicates  $p_{\rm jet}  \lesssim p_{\rm
ism}$ with pressure increasing near the jet edge.  So far, no analytical
or  numerical  solutions, except  $a  = 2$,  have  been  derived in  the
non-magnetized cases.  We note that even in a hydrodynamic recollimation
shock model, the jet boundary  (``shocked zone'' at the interface of the
jet  and the  ISM) is  expected  to be  parabolic. However,  the ISM  is
required to be {\em uniform} ($p_{\rm ism} \sim const. $) \citep[]{KF97,
NS09}.

In  order to  have  parabolic streamlines,  $b  \sim 2$  is required  in
magnetized  jets.   It is  important  to  examine  the distributions  of
$p_{\rm   ism}$.   A   recent  analytical   model  for   a   giant  ADAF
\citep[]{NF11} shows  $b \sim 2.3$ from  beyond $r_{\rm B}$  down to the
SMBH,  and  it is  universally  derived for  slow  rotation  of the  ISM
($\lesssim$ 30  \% of the Keplerian  speed) with the wider  range of the
$\alpha$-viscosity parameter (0.001 - 0.3) \cite[]{SS73}.  A spherically
symmetric accretion  with MHD turbulence \citep[]{S08}  produces $b \sim
2.1$ without a  strong dependence on the parameters  and the equation of
state.  However, the spatial resolution in current X-ray observations is
not high enough to resolve this region.

\subsection{HST-1}

The radius of HST-1 appears to be smaller than the size estimated by the
two  power-law streamlines.  The  maximam measured  radius of  the HST-1
region is 2200 $r_{s}$ $\pm$ 300  $r_{s}$ and mean value is 1300 $r_{s}$
$\pm$ 150 $r_{s}$,  while the expected radii are  3000 $r_{s}$ $\pm$ 400
$r_{s}$ for  the parabolic case and  4100 $r_{s}$ $\pm$  200 $r_{s}$ for
the  conical case,  respectively.  One  possibility is  that the  jet is
subjected to over-collimation towards HST-1.

As is seen  in Fig. \ref{r-z}, HST-1 is located  downstream of the Bondi
radius ($r_{\rm  B}$), where the  jet changes from parabolic  to conical
shape.  This  transition is presumably caused by  the different profiles
of the external pressure (changing from steep to shallow gradients): The
distribution of the  external gas follows the influence  of gravity from
the  central SMBH  inside  of $r_{\rm  B}$,  while it  can be  flattened
outside of $r_{\rm B}$.  The field strength $|\bolB|$ at HST-1 is 0.5 --
20  mG  as  estimated  by  the  X-ray  variability  \citep[]{P03,  H09}.
Supposing $|\bolB| \sim$  1 mG for our reference,  the magnetic pressure
is expected to be about 100 times larger (i.e., over-pressured) than the
ISM  value  $p_{\rm  ism}   \sim  4.3  \times  10^{-10}$  dyn  cm$^{-2}$
\citep[$n_e=0.17$ cm$^{-3}$ and $kT=0.8$ KeV]{A06}.

Another  interesting objective  is understanding  the production  of the
superluminal  components \citep[]{B99,C07}.   The  observed motions  are
modeled  by  the  quad  relativistic  shocks  in  the  MHD  helical  jet
\citep[]{N10}.  Recent  optical polarimetry requires  a coherent helical
field to explain the distribution of polarization vectors (perpendicular
to the  jet) \citep[]{P03} and  variations of the radio  electric vector
positional  angles  as  well   as  the  fractional  polarization  degree
\citep[]{C11}, indicating  either a helical  distortion to the jet  or a
shock propagating through a helical jet \citep[]{P11}.

An   interpretation  for  the   origin  of   HST-1  as   a  hydrodynamic
recollimation  shock has been  proposed by  \citet[][hereafter S06]{S06}
and \citet[][hereafter  BL09]{BL09}.  S06 does not  explain the observed
gradual collimation, and is  inconsistent with our observational result.
They  assume a freely  expanding ({\em  i.e.}, a  conical) jet  which is
already  in  the  particle-dominated  regime  upstream  of  HST-1.   S06
suggests  a  shallow  external  pressure profile  $p_{\rm  ism}  \propto
z^{-0.6}$ inside some critical  radius of 3$\arcsec$ ($\approx$ 234 pc),
as  determined  by  a stellar  cusp  (a  break  in the  stellar  surface
brightness of the  M87 host galaxy) \citep[]{L92}. BL09  also models the
recollimation shock at HST-1 with the same pressure profile.

In  summary, we  propose that  a  transition from  parabolic to  conical
streamlines, occurs via  a strong compression, due to  the change in the
slope  of the axial  profile of  $p_{\rm ism}$.   This condition  can be
achieved by an imbalance between  $p_{\rm jet}$ and $p_{\rm ism}$ in the
radial  direction due to  the over-collimation.   It is  remarkable that
recent General Relativistic  MHD simulations \citep[]{M06} reproduce the
observed transition  from parabolic ($a  \sim 1.7$) to conical  ($a \sim
1.0$) streamlines.  However, the scale  of the transition does not match
with  the observations.   We suggest  that  the properties  of MHD  jets
should be considered in a realistic ISM environment.

\subsection{Towards the Jet Origin: BH or Disk?}

Recent VLBA observations indicate the  VLBI core at 43 GHz (core$_{43}$)
is located at  20 $r_{s}$ from the position of the  SMBH for the viewing
angle of 14$^{\circ}$  \citep[]{H11}.  With the VLBA image,  the size of
the core$_{43}$ can be measured to be 17 $\pm$ 4 $r_{s}$ at the position
angle of  68$^{\circ}$, which is approximately perpendicular  to the jet
axis.  If we assume the core  is the innermost part of the jet emission,
the radius of the  jet is 9 $\pm$ 2 $r_{s}$.  This  is in good agreement
with the  extrapolated radius of the  jet of 9 $\pm$  2 $r_{s}$ assuming
parabolic streamlines in the inner region.  If this is the case, the jet
propagates with a single power-law streamline over more than four orders
of magnitude of distance.

The  jet opening  angle is  one of  the key  parameters in  the  MHD jet
acceleration mechanism.  \citet[][hereafter BP82]{BP82} suggest that the
jet is initiated by the magneto-centrifugal force if the jet inclination
angle between the penetrating magnetic  line of force and accretion disk
is smaller than  60$^{\circ}$, which corresponds to a  jet opening angle
larger  than 60$^{\circ}$.   If we  assume that  the core$_{43}$  is the
innermost part of jet emission, the jet opening angle would be estimated
to be 46$^{\circ}$  at 20 $r_{s}$.  Therefore, the  core$_{43}$ would be
located downstream  of the Alfv\'en surface where  the magnetic pressure
gradient  force plays  a role.   Thus, future  sub-mm VLBI  projects can
access initial acceleration mechanisms by direct imaging and measure the
deviation from parabolic streamlines upstream of the core$_{43}$.

Furthermore,  as we show  in Fig.  \ref{r-z}, the  parabolic jet  can be
extrapolated to the vicinity of  the SMBH.  At the ``axial'' distance of
$z$    =   3.4   $r_{s}$    from   the    SMBH,   it    intersects   the
theoretically-expected,  inner  stable  circular  orbit  (ISCO)  of  the
accretion  disk for  a  Schwarzschild  black hole.   If  the jet  radius
departs from  the extrapolation  line well beyond  3.4 $r_{s}$,  it will
indicate that the  jet originates in the accretion  disk (BP82).  If the
streamline  is  as  extrapolated  down  to 3.4  $r_{s}$,  the  jet  must
originate from  the accretion disk  (BP82) around a spinning  black hole
and/or  the ergosphere  of  the spinning  black hole  \citep[][hereafter
BZ77]{BZ77}.  If we see no departure  down to 0.5 $r_{s}$ from the SMBH,
it indicates that the jet originates from the ergosphere of the spinning
black hole (BZ77).  Note that the above discussion for prograde spinning
black  holes, while  the upper  limit  of the  ISCO is  4.5 $r_{s}$  for
retrograde spinning black holes.   Therefore tracing back the ridge line
and identifying  the ejection  point of  the jet is  one of  the crucial
tests  to an  evaluation of  the formation  mechanism of  the  jet (BP82
and/or BZ77).  More fundamentally, it  provides an important  method for
detection  the spin  of the  SMBH\footnote{It is  also  interesting to
compare  with  the jet  power  as  is  predicted by  \citet[]{G10}}  and
evaluation of  the general theory of relativity.   Such conjectures will
be explored by direct imaging with future submm VLBI experiments.

\acknowledgments

We  thank M. Inoue,  A. Doi,  H. Nagai,  D. Garofalo  and P.T.P.  Ho for
stimulating discussions  and proof reading.  We would  like to sincerely
acknowledge R. Craig Walker for his  providing the VLBA image at 43 GHz.
M.N.  is  grateful  to  D.~L.  Meier  and  C.~A.  Norman  for  inspiring
discussions.   M.N.  was supported  by  the  Allan  C. Davis  fellowship
jointly awarded by JHU and STScI.   The European VLBI Network is a joint
facility of  European, Chinese, South African and  other radio astronomy
institutes  funded by  their national  research councils.   MERLIN  is a
National Facility  operated by the  University of Manchester  at Jodrell
Bank  Observatory  on behalf  of  STFC.   The  National Radio  Astronomy
Observatory is  a facility of  the National Science  Foundation operated
under cooperative agreement by Associated Universities, Inc.

{\it Facilities:} \facility{EVN, MERLIN, VLBA}.

\begin{deluxetable}{lcccccc}
\tabletypesize{\scriptsize}
\tablecaption{Qualities of the images}
\tablewidth{0pt}
\tablehead{
   \colhead{Observation}	& \colhead{}	& \colhead{Synthesised beam}	&  \colhead{}	& \colhead{Peak intensity} & \colhead{RMS noise level} 
   & \colhead{Dynamic Range} \\
   \colhead{} &\colhead{[mas]}	& \colhead{[mas]}	& \colhead{[degree]}	& \colhead{[Jy beam$^{-1}$]} & \colhead{[mJy beam$^{-1}$]} & \colhead{}
   }
\startdata
  EVN         & 17.8	& 13.1	& 76.1$^{\circ}$	& 1.77 & 0.47 & 3770 \\
  MERLIN	& 25	& 25  & 0$^{\circ}$	& 2.00 & 1.21 & 1650 \\
  VLBA	& 1.22	& 0.59	& -3.99$^{\circ}$ 	& 0.838	  & 0.15  & 5590
\enddata
\end{deluxetable}

\end{document}